\documentclass[12pt]{article}
\usepackage{amsmath,amssymb,epsfig,amsfonts}
\usepackage{graphicx,subfigure}
\usepackage[usenames, dvipsnames]{color}
\usepackage[backref]{hyperref}
\usepackage{cite}

\makeatletter

\def\bh{\hat{b}}
\def\ch{\hat{c}}
\def\cb{c_{1,B_3}}
\def\cs{c_{1,S_2}}
\def\alphah{\hat{\alpha}}
\def\eh{\hat{e}}
\def\ah{\hat{a}}

\newcommand{\be}{\begin{equation}}
\newcommand{\ee}{\end{equation}}
\newcommand{\ba}{\begin{aligned}}
\newcommand{\ea}{\end{aligned}}

\def\unit{{1\kern-.65ex {\rm l}}}
\def\1{{1\kern-.65ex {\rm l}}}

\def\CB{{\cal B}}
\def\CC{{\cal C}}

\def\CL{{\cal L}}

\def\CN{{\cal N}}
\def\CO{{\cal O}}

\def\CX{{\cal X}}

\newcount\hour \newcount\minute
\hour=\time \divide \hour by 60
\minute=\time
\def\now{%
\ifnum \hour<13
  \ifnum \hour=0 \advance \hour by 12 \number\hour:\else \number\hour:\fi%
     \ifnum \minute<10 0\fi%
     \number\minute%
\ A.M.%
\else \advance \hour by -12 \number\hour:%
  \ifnum \minute<10 0\fi%
  \number\minute%
  \ P.M.%
\fi%
}

\makeatother

\begin{document}

\baselineskip=18pt  
\numberwithin{equation}{section}  
\allowdisplaybreaks  



%
%


\thispagestyle{empty}

\vspace*{-2cm} 
\begin{flushright}
{\tt EFI-12-27}\\
{\tt KCL-MTH-12-12}\\
\end{flushright}

\vspace*{0.8cm}
\begin{center}
 {\LARGE  Global Gluing and $G$-flux}

 \vspace*{1.5cm}
 Joseph Marsano$^1$, Natalia Saulina$^2$ and Sakura Sch\"afer-Nameki$^3$\\
 \vspace*{1.0cm}
 
{ \it$^1$ Enrico Fermi Institute, University of Chicago\\}
{ \it 5640 S Ellis Avenue, Chicago, IL 60637, USA\\}
{\it and \\}
{\it Sunrise Futures LLC  \\
30 S Wacker Suite 1706, Chicago, IL 60606, USA\\}
{\tt gmail:  joseph.marsano\\}
\smallskip

{\it $^2$ Department of Physics\\}
{\it  Rutgers University, Piscataway, NJ, USA\\}
{\tt saulina theory caltech edu\\}
\smallskip

{ \it $^3$ Department of Mathematics, King's College, University of London \\ }
{ \it The Strand, WC2R 2LS, London, UK\\ }
{\tt  gmail: sakura.schafer.nameki \\}

\vspace*{0.8cm}
\end{center}
\vspace*{.5cm}

\noindent
{Local models that capture the 7-brane physics of F-theory compactifications for supersymmetric GUTs are conveniently described in terms of an $E_8$ gauge theory in the presence of a Higgs bundle.  Though the Higgs bundle data is usually determined by the local geometry and $G$-flux, additional gluing data must be specified whenever the Higgs bundle spectral cover is not smooth.  In this paper, we argue that this additional information is determined by data of the $M$-theory 3-form that is not captured by the cohomology class of the $G$-flux.  More specifically, we show that when the 3-form is specified in terms of a line bundle on a spectral divisor, which is a global extension of the Higgs bundle spectral cover, the gluing data of the local model is uniquely determined in a way that ensures agreement with Heterotic results whenever a Heterotic dual exists.}

\newpage
\setcounter{page}{1} 



\tableofcontents
\newpage

\section{Introduction and Summary}

Though local F-theory models are well-understood by now, recent work has highlighted an interesting subtlety. 
Local F-theory GUTs have largely been described in the framework of the 8-dimensional $\mathcal{N}=1$ SYM theory on the worldvolume of the GUT 7-branes in the presence of a nontrivial Higgs bundle \cite{Donagi:2009ra}.  Though a generic Higgs bundle is uniquely specified by its spectral data, additional data is required when the spectral cover is not smooth \cite{Cecotti:2010bp,Donagi:2011jy,Donagi:2011dv}.  This is particularly relevant when considering models with $U(1)$ symmetries that do not localize along non-Abelian singularities. In these models, the spectral cover $\mathcal{C}_{Higgs}$ splits into multiple components  \cite{Tatar:2009jk, Marsano:2009gv, Marsano:2009wr, Hayashi:2010zp, Marsano:2010ix, Dudas:2010zb, Grimm:2010ez, Marsano:2011nn, Dolan:2011iu} so that the full cover $\mathcal{C}_{Higgs}$ is singular, where the components intersect.  Gluing represents additional data that must be specified along these intersections.


  
The importance of gluing data presents a serious challenge in light of the recent advances in global model building.  Given a global model, we know how to determine the spectral data of the Higgs bundle but what if we need to know more?  How does the global compactification specify the gluing data that completes the specification of the local model?  Addressing this issue in a practical way is the subject of this paper.
 
As gluing data is intrinsically tied to the description of Higgs bundles with spectral covers, we rely heavily on a global generalization of the spectral cover construction.  Any singular elliptic four fold that has a local characterization in terms of a spectral cover $\mathcal{C}_{Higgs}$ in the vicinity of a component of the discriminant exhibits a spectral divisor, $\mathcal{C}_F$, that provides a global extension of that cover \cite{Marsano:2010ix, Marsano:2011nn, MS, Kuntzler:2012bu}\footnote{Other recent progress on global definition of $G$-flux can be found in \cite{Collinucci:2010gz, Braun:2011zm, Krause:2011xj, Krause:2012yh, Collinucci:2012as}.}.  The restriction of $\mathcal{C}_F$ to the (singular elliptic fibration over) the GUT 7-brane locus is a compact version of the Higgs bundle spectral cover $\mathcal{C}_{loc}$ that is isomorphic to the Heterotic spectral cover whenever a Heterotic dual exists \cite{Cvetic:2012ts} {\footnote{In the presence of an additional $U(1)$ symmetry, the divisor $\mathcal{C}_F$ splits into components, thereby extending the reducibility of $\mathcal{C}_{loc}$ in that case.}}.
Just as the gauge bundle of the local model depends on a choice of line bundle on $\mathcal{C}_{Higgs}$, the $G$-flux of the global model can be specified by a choice of line bundle (or sheaf) $\mathcal{N}_F$ on $\mathcal{C}_F$ {\footnote{More specifically, the cohomology class of $G$ is determined only by $c_1(\mathcal{N}_F$).  In principle, the choice of $\mathcal{N}_F$ itself carries more information that is reflected in the 3-form $C_3$.}}.  When the $G$-flux is specified in this way we obtain a unique sheaf by restriction to $\mathcal{C}_{loc}$ with any requisite gluing data completely determined.


The implications  for the spectrum are 
 easily clarified in the most interesting case where the Higgs bundle spectral cover is reducible.  The data that needs to be specified is two-fold
\begin{itemize}
\item[(1)] Do we treat the individual components of $\mathcal{C}_{Higgs}$ separately or not?  Is the gauge bundle determined by separate line bundles on each component or are they glued together?  In the global geometry, the question of whether the bundles are glued or not depends on whether the  spectral divisor $\mathcal{C}_F$ is factored or not.  This impacts the map that determines the cohomology class of $G$ in terms of a divisor on $\mathcal{C}_F$ \cite{Cvetic:2012ts} and can thereby influence the chirality of the spectrum.

\item[(2)] If the Higgs bundle spectral cover splits into components whose bundles are glued then how are they glued?  This amounts to specifying a map from the total space of the line bundle on one component to the total space of the line bundle on another component when restricted to the intersection locus \cite{Donagi:2011jy,Donagi:2011dv}.  From the local model perspective, this represents new data that must be specified in addition to the spectral data.  We argue that it is determined by information about $C_3$, that is not necessarily captured by the cohomology class of $G$.  As expected, such data does not change the net chirality but does impact the precise spectrum. 

 
\end{itemize}

The effect on the spectrum is easy to see.  Consider the {spectrum} computation in the context of  a factored spectral cover $\mathcal{C}_{Higgs} = \mathcal{C}_{Higgs}^{(a)} \mathcal{C}_{Higgs}^{(b)}$ in an $SU(5)_{\rm GUT}$ model.   
The $\mathbf{10}$ matter curve in the compactification of the spectral cover, $\mathcal{C}_{loc}$, splits into two components, $\Sigma^{(a)}_{\bf 10}$  and $\Sigma^{(b)}_{\bf 10}$. The spectrum is determined by 
\begin{equation}\label{Chirality}
 H^m(\Sigma_{\mathbf{10}},{\CL_{\mathbf{10}}})\,, \quad {\CL_{\mathbf{10}}}:=p_{loc}^*K_{S_2}\otimes \CN_{loc}\vert_{\Sigma_{10}}\,,
\end{equation}
which on the other hand can be computed from the exact sequence \cite{Donagi:2011jy, Donagi:2011dv}
\begin{equation}\begin{split}\label{TheSeq}
0 &\rightarrow H^0(\Sigma_{\mathbf{10}}^{(b)},\CL_{\mathbf{10}}(-\Sigma_{\mathbf{10}}^{(a)})|_{\Sigma_{\mathbf{10}}^{(b)}}) \stackrel{\iota}{\rightarrow} {H^0(\Sigma_{\mathbf{10}},\CL_{\mathbf{10}})}\rightarrow H^0(\Sigma_{\mathbf{10}}^{(a)},\CL_{\mathbf{10}}|_{\Sigma_{\mathbf{10}}^{(a)}}) \\
&\stackrel{\partial}{\rightarrow}  H^1(\Sigma_{\mathbf{10}}^{(b)},\CL_{\mathbf{10}}(-\Sigma_{\mathbf{10}}^{(a)})|_{\Sigma_{\mathbf{10}}^{(b)}}) \rightarrow {H^1(\Sigma_{\mathbf{10}},\CL_{\mathbf{10}})}\rightarrow H^1(\Sigma_{\mathbf{10}}^{(a)},\CL_{\mathbf{10}}|_{\Sigma_{\mathbf{10}}^{(a)}}) \rightarrow 0 \,.
\end{split}\end{equation}
Here $H^0(\Sigma_{\mathbf{10}}^{(b)},\CL_{\mathbf{10}}(-\Sigma_{\mathbf{10}}^{(a)})|_{\Sigma_{\mathbf{10}}^{(b)}})$ counts sections of 
$\CL_{\mathbf{10}}|_{\Sigma_{\mathbf{10}}^{(b)}}$ that vanish at intersection points $\Sigma_{\mathbf{10}}^{(a)}\cap \Sigma_{\mathbf{10}}^{(b)}.$ The map 
$\partial$ is the coboundary map which depends on the gluing morphism
\cite{Donagi:2011jy, Donagi:2011dv}.

If $\partial$ is trivial, then the sequence splits and (\ref{TheSeq}) implies\footnote{If we write $\CL_{\mathbf{10}}=\CO_{\Sigma_{\mathbf{10}}}(\gamma_a +\gamma_b)\otimes K_{\Sigma_{\mathbf{10}}}^{1/2},$ then the chirality of the spectrum in (1.3) agrees with a sum of two independent chiralities $\sum_{m=0}^1 (-1)^m h^m\Bigl(\Sigma_{\mathbf{10}}^{(a)},
\CO_{\Sigma_{\mathbf{10}}^{(a)}}(\gamma_a)\otimes K_{\Sigma_{\mathbf{10}}^{(a)}}^{1/2}\Bigr)$ and $\sum_{m=0}^1 (-1)^m h^m\Bigl(\Sigma_{\mathbf{10}}^{(b)},
\CO_{\Sigma_{\mathbf{10}}^{(b)}}(\gamma_b)\otimes K_{\Sigma_{\mathbf{10}}^{(b)}}^{1/2}\Bigr)$ as appropriate
for the case with trivial gluing morphism.}
\begin{equation}\label{spec}
 h^m(\Sigma_{\mathbf{10}},\CL_{\mathbf{10}}) =h^m(\Sigma_{\mathbf{10}}^{(b)},\CL_{\mathbf{10}}(-\Sigma_{\mathbf{10}}^{(a)})|_{\Sigma_{\mathbf{10}}^{(b)}}) +
 h^m\left(\Sigma_{\mathbf{10}}^{(a)},\CL_{\mathbf{10}}|_{\Sigma_{\mathbf{10}}^{(a)}}\right) \,.
\end{equation}
Note that the chirality of the spectrum in (\ref{spec}) differs from the chirality
of the spectrum for an irreducible $\Sigma_{\mathbf{10}}.$ From the global perspective
this  is quite natural. As we recall in (3.6) below, the way G-flux is read off from the bundle $\mathcal{N}_F$
depends on the normal bundle to $\mathcal{C}_{loc}$ inside the spectral divisor $\mathcal{C}_F,$ and this normal bundle changes when  $\mathcal{C}_F$ is deformed
such that $\mathcal{C}_{loc}$ becomes reducible \cite{Cvetic:2012ts}.  The presence of gluing data can affect the details of $\partial$ but, as is clear from \eqref{spec}, this does not change the net chirality of the spectrum.


We will first briefly review the local models in F-theory and gluing. In section 3, we introduce the compactification $\mathcal{C}_{loc}$ of $\mathcal{C}_{Higgs}$ and explain how a line bundle on the spectral divisor gives rise to gluing in the local limit. This is exemplified in section 3. Finally, in section 4 we conjecture what happens when the global spectral divisor factors: there is the possibility that the spectral divisor splits, indicating the presence of an extra $U(1)$. This setup allows specification of additional data, globally, which we will conjecture to correspond to a non-abelian potential $C_3$.


\section{Local Model and Gluing}

Consider F-theory on a singular elliptic Calabi-Yau fourfold, and let $S_2$ be a component of the discriminant, corresponding to 7-branes wrapping $S_2\times \mathbb{R}^{3,1}$. For an $A_{N-1}$ type singularity the low energy effective  theory on the  7-brane is described by an 8-dimensional $\mathcal{N}=1$ SYM. We begin by describing what determins this local model, i.e. the data specifying the SYM theory. 

\subsection{Higgs bundle and spectral cover $\CC_{Higgs}$}

The holomorphic data of a typical local model is comprised of an $SU(N)$ Higgs bundle on a compact surface $S_2$.  This Higgs bundle breaks an underlying $E_8$ Yang-Mills theory on $\mathbb{R}^{3,1}\times S_2$ to a phenomenologically interesting subgroup, such as $SU(5)$ or $SO(10)$, and is typically specified by a spectral cover $\CC_{Higgs}$
\begin{equation}p_{{Higgs}}:\CC_{Higgs}\rightarrow S_2\end{equation}
along with a line bundle $\CN_{Higgs}$ on $\CC_{Higgs}$.  The bundle $\CN_{Higgs}$ determines a non-Abelian vector bundle $V$ on $S_2$ via
\begin{equation}V = p_{{Higgs},*}\CN_{Higgs}\end{equation}
and the condition that it be an $SU(N)$ bundle, $c_1(V)=0$, motivates a decomposition of $\CN_{Higgs}$ according to
\begin{equation}c_1(\CN_{Higgs})=\gamma_{Higgs}+\frac{r_{Higgs}}{2}\,,
\end{equation}
where $r_{Higgs}$ is the ramification divisor of $p_{{Higgs}}$ and the pushforward of $\gamma_{Higgs}$ as a divisor is trivial
\begin{equation}p_{{Higgs},*}\gamma_{Higgs}=0 \,.
\label{tracelessgamma}\end{equation}

In a global model, where $S_2$ is a component of the discriminant locus of some elliptically fibered Calabi-Yau 4-fold $Y_4$, $\CC_{Higgs}$ is determined by the local geometry near $S_2$, while the bundle data is determined by the M-theory {4-form flux $G_4$}.  This dictionary has been studied in many places, initially in \cite{Donagi:2009ra} and has been made quite precise using the formalism of spectral divisors \cite{Marsano:2010ix, Marsano:2011nn, MS, Kuntzler:2012bu}. 

 To review the
 subtlety, let us work with the particular example of an $SU(5)$ model where $\CC_{Higgs}$ is a hypersurface inside the total space of the canonical bundle $K_{S_2}$ that covers $S_2$ 5 times
\begin{equation}
\CC_{Higgs}:\quad b_0 s^5 + b_2 s^3 + b_3 s^2 + b_4 s + b_5=0\,,
\label{CHiggs}\end{equation}
where the objects appearing are sections of the indicated bundles on $S_2$
\begin{equation}\begin{array}{c|c}
\text{Section} & \text{Bundle} \\ \hline
s & c_1 \\
b_m & \CO(\eta-mc_1)
\end{array}\end{equation}
Here, $c_1$ is shorthand for an anti-canonical divisor on $S_2$ and $\eta$ is a divisor class that is determined by $c_1$ and the normal bundle of $S_2$ in $B_3$, $N_{S_2/B_3}=\CO(-t)$, in the usual way
\begin{equation}\eta = 6c_1-t\,.
\end{equation}
The local geometry of a global model determines the $b_m$'s while the $G$-flux determines $\gamma_{Higgs}$ and consequently $\CN_{Higgs}$ from
\begin{equation}\CN_{Higgs} = \CO_{\CC_{Higgs}}(\gamma_{Higgs}+r_{Higgs}/2)\,,
\end{equation}
where $\gamma_{Higgs}+r_{Higgs}/2$ is guaranteed to be an integral class when the $G$-flux is properly quantized.


\subsection{{Gluing in local models}}

The data described so far is however not complete, in particular when $\CC_{Higgs}$ is not smooth. 
As a simple example of this type of situation, suppose $\CC_{Higgs}$ splits into two components as in
\begin{equation}
\label{FactoredHiggs}
\CC_{Higgs} = \CC_{Higgs}^{(a)}\times \CC_{Higgs}^{(b)}  \,.
\end{equation}
The $G$-flux determines a divisor $\gamma_{Higgs}$ that will have restrictions to each of the two components, $\CC_{Higgs}^{(a)}$ and $\CC_{Higgs}^{(b)}$, respectively
\begin{equation}
\gamma_{Higgs}^{(a)}\equiv \gamma_{Higgs}|_{\CC_{Higgs}^{(a)}} \,,
\qquad \gamma_{Higgs}^{(b)}\equiv \gamma_{Higgs}|_{\CC_{Higgs}^{(b)}} \,.
\end{equation}
Each of these, in turn, can be used to determine a line bundle on the corresponding component
\begin{equation}\CN_{Higgs}^{(m)} 
= \CO_{\CC_{Higgs}^{(m)}}\left(\gamma_{Higgs}^{(m)}+\frac{r_{Higgs}^{(m)}}{2}\right)\,,
\end{equation}
where $r_{Higgs}^{(m)}$ is the ramification divisor of $\CC_{Higgs}^{(m)}$.  This is not enough to fully specify $\CN_{Higgs}$, however, because { in local models there is no unique prescription} what to do at the singular points of $\CC_{Higgs}$ where the two individual components meet.  The extra data that must be specified here is the gluing data \cite{Donagi:2011jy, Donagi:2011dv} and is related to a type of `T-brane configuration' \cite{Cecotti:2010bp}. 


The physics is sensitive to $\CN_{Higgs}$ primarily through its restriction to various matter curves as we obtain, in this way, the line bundle to which physical degrees of freedom couple.  
Consider for instance the $\mathbf{10}$ matter curve $\Sigma_{\mathbf{10}}$ inside $\CC_{Higgs}$
\begin{equation}\Sigma_{\mathbf{10}}:\quad b_5=s=0 \,.
\end{equation}
The degrees of freedom that localize there are counted by the standard cohomologies
\begin{equation}
H^p(\Sigma_{\mathbf{10}},p_{Higgs}^*K_{S_2}\otimes \CN_{Higgs}|_{\Sigma_{\mathbf{10}}}) \,.
\label{tencohoms_i}\end{equation}
When $\CC_{Higgs}$ splits into components as in (\ref{FactoredHiggs}), the $\mathbf{10}$ matter curve itself becomes reducible
\begin{equation}
\Sigma_{\mathbf{10}} = \Sigma_{\mathbf{10}}^{(a)}
\cup \Sigma_{\mathbf{10}}^{(b)} \,.
\end{equation}
The nature of $\CN_{Higgs}|_{\Sigma_{\mathbf{10}}}$ depends crucially on how $\CN_{Higgs}$ behaves along the locus where the two components, $\CC_{Higgs}^{(a)}$ and $\CC_{Higgs}^{(b)}$, meet.  
The computation of the cohomologies (\ref{tencohoms_i}) is not necessarily given by 
a sum
\begin{equation}
H^p(\Sigma^{(a)}_{\mathbf{10}},p_{Higgs}^*K_{S_2}\otimes \CN_{Higgs}|_{\Sigma^{(a)}_{\mathbf{10}}}) \oplus H^p(\Sigma^{(b)}_{\mathbf{10}},p_{Higgs}^*K_{S_2}\otimes \CN_{Higgs}|_{\Sigma^{(b)}_{\mathbf{10}}}) \,.
\end{equation}
It is natural to then ask: if local geometry determines the $b_m$'s and $G$-flux determines $\gamma_{Higgs}$, to what does the gluing data correspond to globally?  Said differently, given a Calabi-Yau fourfold $Y_4$ with $SU(5)$ singularity along $S_2$, how do we determine the gluing data of the resulting local model? We will address this in the next section.


\section{Global Gluing}

As the gluing is formulated in the context of the Higgs bundle spectral cover description of the local model, we should work in a formalism where the map from global data to local spectral covers is understood.  In fact, for any compact CY fourfold $Y_4$ with an ADE type singularity over a surface $S_2$, we will define a compact version $\mathcal{C}_{loc}$ of the Higgs bundle spectral cover \cite{Cvetic2012ts}, which is a restriction of the globally defined, compact spectral divisor $\mathcal{C}_F$. $\mathcal{C}_{loc}$ can be defined more succinctly in terms of an auxiliary three-fold $Z_3$. This agrees with the heterotic CY threefold whenever a heterotic dual exists, but crucially can be defined in general. 

From the global point of view, gluing is encoded in a line bundle $\mathcal{N}_F$ over the spectral divisor $\mathcal{C}_{F}$, which restricts to a bundle on $\mathcal{C}_{loc}$. This is globally an entirely well-defined object, irrespective of whether $\mathcal{C}_{Higgs}$ is smooth or not. The Higgs bundle interpretation depends on those aspects, in particular for smooth $\mathcal{C}_{Higgs}$ the restriction of $\mathcal{N}_F$ yields a line bundle on $\mathcal{C}_{Higgs}$, which  corresponds in the usual way to local flux. If the Higgs bundle spectral cover is not smooth, generically we obtain a sheaf on $\mathcal{C}_{Higgs}$, which we can interpret as local flux plus gluing information. Again, from the global perspective there is no distinction between these for a given line bundle $\mathcal{N}_F$ on the spectral divisor. 

We will now explain these points by first revisiting the concept of spectral divisors in the global model, which define by restriction the compactification $\mathcal{C}_{loc}$ of the Higgs bundle spectral cover, and finally, we describe  gluing in this context.


\subsection{Spectral Divisor $\CC_F$}

We stick to $SU(5)$ models for concreteness and look, then, to a Calabi-Yau 4-fold $Y_4$ defined as the following hypersurface inside a $\mathbb{P}^2_{1,2,3}$ bundle over a base $B_3$
\begin{equation}
y^2 = x^3 + v\left[\CB_0 z^5 + \CB_2 z^3v^3 x + \CB_3 z^2v^2 y + \CB_4 zvx^2 + \CB_5xy\right] \,.
\end{equation}
In general, $\CB_m$ is not identically zero at $z=0$.  It may contain a piece that is, however.  Let us remove this ambiguity and write, in general
\begin{equation}\CB_m = \bh_m + \ch_m z \,,
\end{equation}
where we take $\bh_m$ so that $\bh_m = \CB_m(z=0)$.  In this way, we arrive at $Y_4$ of the form
\begin{equation}\begin{split}
y^2 &= x^3 + \ch_0 (zv)^6 + \ch_2 (zv)^4 x + \ch_3 (zv)^3 y + \ch_4 (zv)^2x^2 + \ch_5 xy (zv) \\
& + v\left[\bh_0 (zv)^5 + \bh_2 (zv)^3 x + \bh_3 (zv)^2 y + \bh_4 (zv) x^2 + \bh_5 xy\right] \,,
\label{Y4def}
\end{split}\end{equation}
where the objects here are sections of the indicated bundles
\begin{equation}\begin{array}{c|c}
\text{Section} & \text{Bundle} \\ \hline
v & \CO(\sigma) \\
x & \CO(2[\sigma+\cb]) \\
y & \CO(3[\sigma + \cb]) \\
\bh_m & \CO([6-m]\cb - [5-m]S_2) \\
\ch_m & \CO([6-m](\cb-S_2))
\end{array}\end{equation}
with $\sigma$ the section of $Y_4$ and $\cb$ shorthand for an anti-canonical divisor of $B_3$.  
As the $\ch_m$'s represent 'higher order behavior', they can be modified or even completely removed by making replacements of the form $\bh_m\rightarrow \bh_m - c'_m z$.  The condition $\bh_m = \CB_m(z=0)$ completely fixes any higher order ambiguity and we will assume it throughout.

Given $Y_4$, we define the spectral divisor $\CC_F$ as the 5-sheeted cover of $B_3$ that emerges from the proper transform of the 3-fold
\begin{equation}
\mathcal{C}_F:\qquad 
\ba
y^2 &= x^3 + \ch_0(zv)^6 + \ch_2(zv)^4 x + \ch_3(zv)^3 y + \ch_4 (zv)^2 x^2 + \ch_5 (zv)xy \cr
0 &= \bh_0 z^5 + \bh_2 z^3 x + \bh_3 z^2 y + \bh_4 zx^2 + \bh_5 xy
\ea\end{equation}
in the Calabi-Yau resolution $\tilde{Y}_4$ of $Y_4$.  The usefulness of $\CC_F$ is that it provides a bridge that connects $G$-flux to local model data.  
We can specify $G$-fluxes in $Y_4$, for instance, by choosing a line bundle $\CN_F$ on $\CC_F$ and forming
\begin{equation}G = \iota_{F,*}\left[c_1(\CN_F) - \frac{1}{2}c_1(\CL_{\hat{r}})\right]\end{equation}
where 
\begin{equation}
\iota_F:\quad \CC_F\rightarrow \tilde{Y}_4
\end{equation}
is the embedding map and $\CL_{\hat{r}}$ is a line bundle on $\CC_F$ that is related to the ramification divisor of the covering \cite{Cvetic:2012ts} {\footnote{More specifically, $\CL_{\hat{r}}$ is a twist of the bundle $\CO_{\CC_F}(r_F)$ where $r_F$ is the ramification divisor of $p_F$.  This twist is obtained by extending the normal bundle of $\CC_{loc}$ in $\CC_F$ to a bundle on all of $\CC_F$ and removing the 'vertical pieces', that is the pieces pulled back from $B_3$ via $p_{{F}}$.  The result is a bundle $\CL_{\hat{r}}$ such that $\CL_{\hat{r}}|_{\CC_{loc}}=\CO_{\CC_{loc}}(r_{loc})$.}}
\begin{equation}
 p_F:\quad \CC_F\rightarrow B_3 \,.
 \end{equation}


\subsection{Compact spectral cover $\CC_{loc}$}
\label{subsec:Cloc}

Each singular CY fourfold contains a compactification of the Higgs bundle spectral cover. The restriction of  $\CC_F$ to the singular elliptic fibration over $S_2$ is a compactification of $\CC_{Higgs}$ denoted by $\CC_{loc}$ \cite{Cvetic:2012ts}
\begin{equation}
\CC_{loc}\equiv \CC_F|_{\pi^*S_2} \,,
\label{cclocint}\end{equation}
where 
\begin{equation}
\pi:\quad Y_4 \rightarrow B_3 \,.
\end{equation}
By studying a resolution $\tilde{Y}_4$ of $Y_4$, one finds a simple presentation of the surface $\CC_{loc}$ as a complete intersection in an $\mathbb{F}_1$-fibration over $S_2$ \cite{Cvetic:2012ts}.  More specifically, consider an $\mathbb{F}_1$ surface with homogeneous coordinates $[X,W]$ on the base and $[u,q]$ on the fiber.  We fiber this over $S_2$ to obtain a 4-fold $E_4$ 
\begin{equation}\begin{array}{cc}
\mathbb{F}_1\  \rightarrow &E_4 \cr
&   \downarrow  \cr
& S_2 
\end{array}\end{equation}
in which $X,W,u,q$ transform as sections of the indicated bundles
\begin{equation}\begin{array}{c|c}
\text{Section} & \text{Bundle} \\ \hline
X & \CO(f + 2 c_1) \\
W & \CO(f) \\
u & \CO(b + f + 3c_1) \\
q & \CO(b)
\end{array}\end{equation}
where we introduced notation $b$ and $f$ for the divisors associated to the base and fiber of the $\mathbb{F}_1$.  The intersection 
\eqref{cclocint} is isomorphic to the surface defined by
\begin{equation}
\CC_{loc}: \qquad \ba Wq^2 &= u\left[uX^3 + c_0 uW^3 + c_2uW^2X + c_3 qW^2 + c_4uWX^2 + c_5 qWX\right] \\ 
0 &= b_5 Xq + b_4 uX^2 + b_3Wq + b_2 u WX + b_0 W^2 u \,,
\ea
\end{equation}
where
\begin{equation}
b_m = \bh_m|_{S_2}\qquad c_m = \ch_m|_{S_2} \,.
\end{equation}
The first equation defines an elliptically fibered Calabi-Yau 3-fold $Z_3$ over $S_2$ 
\begin{equation}\begin{array}{cc}
\mathbb{E}\  \rightarrow &Z_3 \cr
&   \downarrow  \cr
& S_2 
\end{array}\end{equation}
while the second gives the 5-sheeted cover $\CC_{loc}$ of $S_2$.  The section of $Z_3$ is along $W=u=0$ and the geometry near this section is nothing other than $\CC_{Higgs}$ \eqref{CHiggs}. {It is worthwhile noting that $Z_3$ agrees with the heterotic CY, whenever a heterotic dual exists. The construction presented here applies to all CY fourfolds,  which makes this approach using $\CC_{loc}$ very powerful.}

 It is helpful to have a more standard description of the auxiliary 3-fold $Z_3$ as a hypersurface in a $\mathbb{P}^2_{1,2,3}$ bundle with weighted homogeneous coordinates $[v,x,y]$.  Denote this bundle over $S_2$ by $\CX_4$.
First  note that $W=0$ everywhere on $Z_3$ where $u=0$.  This motivates the replacement
\begin{equation}W\rightarrow V^2\qquad X\rightarrow X\qquad q\rightarrow Y\qquad u\rightarrow V\,.
\label{replacement}\end{equation}
We obtain an equivalent description of $Z_3$ as the hypersurface
\begin{equation}Y^2 = X^3 + c_0 V^6 + c_2 V^4 X + c_3 V^3 Y + c_4 V^2X^2 + c_5 VXY
\end{equation}
and of $\CC_{loc}$ as the cover specified by
\begin{equation}0=b_5 XY + b_4 X^2V + b_3 YV^2 + b_2 XV^3 + b_0V^5 \,.
\label{Clochet}\end{equation}


It is now easy to see how the choice of line bundle $\CN_F$ specifies the line bundle $\CN_{Higgs}$ of the local model.  We simply restrict $\CN_F$ to $\CC_{loc}$ to obtain a line bundle $\CN_{loc}$ that determines a line bundle $\CN_{Higgs}$ on $\CC_{Higgs}$ when we zoom in near the section
\begin{equation}\CN_F|_{\CC_{loc}} = \CN_{loc}\rightarrow \CN_{Higgs} \,.
\end{equation}
As long as $\CN_F$ is well-defined, the result of this prescription is completely unambiguous whether $\CC_{Higgs}$ is smooth or not.  If it is smooth, we get an ordinary line bundle $\CN_{Higgs}$ on $\CC_{Higgs}$.  If not, we get a sheaf that may admit a description involving a combination of line bundles and gluing data.  Either way, the precise nature of $\CN_{Higgs}$ is completely fixed by the global data, that is the choice of $\CN_F$.
Note that this requires a choice of line bundle on $\CC_F$, which determines the 3-form $C_3$.  In principle this carries more information than the first Chern class, $c_1(\CN_F)$, which determines the flux $G_4$.

Note that it is trivial to specialize to $SO(10)$ or $E_6$ singularities by switching off the relevant $\bh_m$'s and $\ch_m$'s.  It should also be straight forward to construct $SU(6)$ examples  by imposing the right factoring rule on $\CC_{loc}$.


\subsection{Gluing} 

In general, both $\CC_F$ and $\CC_{loc}$ are irreducible (and further $\CC_F$ is smooth)  even when $\CC_{Higgs}$ is not.  As an example, suppose the $\bh_m$'s in \eqref{Y4def} can be written as
\begin{equation}\begin{split}
\bh_0 &= - \alphah\eh_0^2 \\
\bh_2 &= \ah_2\eh_0 + \eh_1^2\alphah \\
\bh_3 &= \ah_3\eh_0 + \ah_2\eh_1 \\
\bh_4 &= \ah_4\eh_0 + \ah_3 \eh_1 \\
\bh_5 &= \ah_4\eh_1 \,.
\end{split}\label{bhfactor}\end{equation}
In this case, $\CC_{Higgs}$ splits into components 
\be
\CC_{Higgs} = \CC_{Higgs}^{(4)} \times \CC_{Higgs}^{(1)}= \left(a_4 + a_3 s + a_2 s^2 + \alpha \left[1 - e_0s\right]\right)(1+e_0s) \,,
\label{CHiggssplits} 
\ee
with
\begin{equation}
\alpha=\alphah|_{S_2}\qquad a_m = \ah_m|_{S_2}\qquad e_n = \eh_n|_{S_2} \,.
\end{equation}
To simplify calculations, the local model literature often compactifies $\CC_{Higgs}$ by compactifying the ambient space, $K_{S_2}$, to $\mathbb{P}(\CO_{S_2}\oplus K_{S_2})$.   The resulting compactification of $\CC_{Higgs}$ preserves its factorization structure.  This obscures the fact, however, that the actual surface $\CC_{loc}$ \eqref{cclocint} that arises from the global geometry is irreducible for generic $\ch_m$.  Correspondingly, the spectral divisor $C_F$ is also irreducible and, moreover, it is generically smooth even when $\CC_{loc}$ is not.

In the end, we have a prescription for determining the local model bundle data by simply restricting the bundle $\CN_F$ on the smooth divisor $\CC_F$ to $\CC_{loc}$ and finally to $\CC_{Higgs}$.  
We will now comment on the consistency of this prescription.

As we have seen in the section on local gluing,  the chiralities of matter in a split spectral cover will crucially depend on the gluing data. In the local model, they are computed by the cohomologies
\begin{equation}H^p(\Sigma_{\mathbf{10}},p_{Higgs}^*K_{S_2}\otimes \CN_{Higgs}|_{\Sigma_{\mathbf{10}}}) \,.
\label{tencohoms}\end{equation}
It is now clear how the restriction of the bundle $\mathcal{N}_F$ can impact these:
The prescription of restricting the bundle $\CN_F$ on $\CC_F$ to $\CC_{loc}$, and ultimately to $\Sigma_{\mathbf{10}}$ and the other matter curves, determines the chiralities.  In fact, we can compute the correct cohomologies \eqref{tencohoms} uniquely by realizing $\Sigma_{\mathbf{10}}$ as a complete intersection inside the smooth space $\CC_F$ and using Koszul sequences.  Often, the restriction $\CN_{loc}$ of $\CN_F$ can alternatively be written as the restriction of a line bundle $\CN_Z$ on the auxiliary 3-fold $Z_3$ to $\CC_{loc}$ in which case Koszul sequences involving the simpler ambient space $Z_3$ can be used.
From this point of view the spectrum of {\bf 10} matter is given by 
\begin{equation}
H^m(\Sigma_{\mathbf{10}},p_{loc}^*K_{S_2}\otimes \CN_{loc}) = H^m(\Sigma_{\mathbf{10}},\CL_{\mathbf{10}})\,,
\end{equation}
where ${\Sigma_{\bf 10}}$ is the matter curve in $\mathcal{C}_{loc}$, and 
$\mathcal{N}_{loc}$ is the restriction of $\mathcal{N}_F$ to $\mathcal{C}_{loc}$. 
These comologies can then be computed from
\begin{equation}\begin{split}
0 &\rightarrow H^0(\Sigma_{\mathbf{10}}^{(b)},\CL_{\mathbf{10}}(-\Sigma_{\mathbf{10}}^{(a)})|_{\Sigma_{\mathbf{10}}^{(b)}}) \rightarrow {H^0(\Sigma_{\mathbf{10}},\CL_{\mathbf{10}})}\rightarrow H^0(\Sigma_{\mathbf{10}}^{(a)},\CL_{\mathbf{10}}|_{\Sigma_{\mathbf{10}}^{(a)}}) \\
&\rightarrow  H^1(\Sigma_{\mathbf{10}}^{(b)},\CL_{\mathbf{10}}(-\Sigma_{\mathbf{10}}^{(a)})|_{\Sigma_{\mathbf{10}}^{(b)}}) \rightarrow {H^1(\Sigma_{\mathbf{10}},\CL_{\mathbf{10}})}
\rightarrow H^1(\Sigma_{\mathbf{10}}^{(a)},\CL_{\mathbf{10}}|_{\Sigma_{\mathbf{10}}^{(a)}}) \rightarrow 0 \,.
\end{split}\label{seq1}\end{equation}
This procedure gives the right sheaf $\CN_{Higgs}$ on $\CC_{Higgs}$ and, correspondingly, the right restrictions to all matter curves because it agrees with the Heterotic side whenever a Heterotic dual exists.  This is almost immediate from our formalism because the 3-fold $Z_3$ is isomorphic to the Heterotic Calabi-Yau in these cases and $\CC_{loc}$ is isomorphic to the Heterotic spectral cover.  The bundle $\CN_{loc}$, then, is isomorphic to the Heterotic spectral bundle $\CN_{Het}$ so the restrictions to various matter curves are guaranteed to agree on both sides.




\section{Example}

We now consider an example, where we compute the precise spectrum in a global model where the local model spectral cover splits. 
There are plenty of examples in \cite{Donagi:2011jy, Donagi:2011dv} for the case when the coboundary map in (\ref{TheSeq}) is non-trivial, i.e. the second case outlined in the introduction.
For this reason we focus on an example of the first type, where the overall chirality changes compared to the case without gluing. 


\subsection{The Setup}

Consider for the base of the elliptic fibration $B_3=\mathbb{P}^3$ and $S_2=H$ a hyperplane in $\mathbb{P}^3$.  The $\bh_m$ and $\ch_m$ are sections of
\begin{equation}\begin{array}{c|c}
\text{Section} & \text{Bundle} \\ \hline
\bh_m & (19-3m)H \\
\ch_m & (18-3m)H
\end{array}\end{equation}

The data relevant for the local model is
\begin{equation}
\eta = 19H\,,\qquad t = -H\,,
\end{equation}
so that we can build an example with the same normal bundle but with a Heterotic dual by replacing the 3-fold base by $\mathbb{P}(\CO_{\mathbb{P}^2}\oplus \CO_{\mathbb{P}^2}(1))$.  Let us stick with $\mathbb{P}^3$ for now.

The local model spectral cover $\CC_{loc}$ is a hypersurface inside the auxiliary 3-fold $Z_3$
\begin{equation}Y^2 = X^3 + c_0 V^6 + c_2 V^4 X + c_3 V^3 Y + c_4 V^2 X^2 + c_5 XYV \,,
\end{equation}
where $V,X,Y$ are projective coordinates on the $\mathbb{P}^2_{1,2,3}$ fiber of the ambient space and transform as sections of
\begin{equation}\begin{array}{c|c}
\text{Section} & \text{Bundle} \\ \hline
V & \CO(\sigma_{loc}) \\
X & \CO(2\sigma_{loc}+6H) \\
Y & \CO(3\sigma_{loc}+9H)
\end{array}\end{equation}
Inside the $\mathbb{P}^2_{1,2,3}$ bundle over $S_2$, which we call $\CX_4$ for definiteness, $Z_3$ is in the class
\begin{equation}Z_3 = 6\sigma_{loc} + 18 H \,.
\end{equation}
The defining equation of the spectral cover is
\begin{equation}\CC_{loc}:\quad 0 = b_0V^5 + b_2V^3X + b_3 V^2Y + b_4VX^2 + b_5XY\end{equation}
and is in the class
\begin{equation}\CC_{loc}:\quad 5\sigma_{loc} + 19H\end{equation}
The traceless $\gamma_{loc}$ is
\begin{equation}\gamma_{loc} = 5\sigma_{loc} - (\eta-5\cs) = 5\sigma - 4H\end{equation}
while the ramification divisor $r_{loc}$ is
\begin{equation}
r_{loc} = \cs + \CC_{loc} = 5\sigma_{loc}+22H \,.
\end{equation}
The inherited bundle, then, is
\begin{equation}\begin{split}\CN_{loc} &= \CO_{\CX_4}\left(\lambda[5\sigma_{loc} - 4H] + \frac{1}{2}[5\sigma_{loc} + 22H]\right)|_{\CC_{loc}} \\
&= \CO_{\CX_4}\left(5\left[\lambda + \frac{1}{2}\right]\sigma_{loc} + \left[11-4\lambda\right]H\right)|_{\CC_{loc}}
\end{split}\end{equation}
and it is by now well-known how to construct a bundle $\CN_F$ on $\CC_F$ that restricts to $\CN_{loc}$.  Let us consider the spectrum of $\mathbf{10}$'s, which is determined by
\begin{equation}H^m(\Sigma_{\mathbf{10}},p_{loc}^*K_{S_2}\otimes \CN_{loc}) = H^m(\Sigma_{\mathbf{10}},\CL_{\mathbf{10}})\,,\end{equation}
where
\begin{equation}\CL_{\mathbf{10}} =  \CO_{\CX_4}\left(5\left[\lambda+\frac{1}{2}\right]\sigma_{loc}+\left[8-4\lambda\right]H\right)|_{\Sigma_{\mathbf{10}}}\,,
\end{equation}
where $\Sigma_{\mathbf{10}}$ is the $\mathbf{10}$ matter curve in $\CC_{loc}$ defined by the restriction of
\begin{equation}b_5=0\end{equation}
to $\CC_{loc}$.  It is isomorphic to the curve $b_5=0$ inside $S_2$.  Since $b_5$ is a section of $\CO_{\mathbb{P}^2}(4H)$ it is a curve of genus 3 and
\begin{equation}\chi(\Sigma_{\mathbf{10}},\CL) = \text{deg}\CL + (1-g) =\text{deg}\CL - 2\end{equation}
As a quick check note that
\begin{equation}\text{deg}\CL_{\mathbf{10}} = 2 - 76\lambda \,.
\end{equation}
Since
\begin{equation}\eta\cdot_{S_2}(\eta-5\cs) = 19H\cdot_{\mathbb{P}^2} 4H = 76\end{equation}
this is consistent with the fact that we know the net chirality should be
\begin{equation}-\lambda\eta\cdot_{S_2}(\eta-5\cs) \,.
\end{equation}
So this indeed defines the line bundle $\CL_{\mathbf{10}}$ correctly. 


\subsection{Chirality computation in $\CC_{loc}$}

 Since the bundle $\CL_{\mathbf{10}}$ is inherited from $\CX_4$ and $\Sigma_{\mathbf{10}}$ is a complete intersection in $\CX_4$, we can use Koszul techniques as in \cite{Blumenhagen:2010pv, cohomCalg:Implementation} to compute the desired cohomologies.  We can describe $\CX_4$ as a toric variety with the GLSM matrix
\begin{equation}\begin{pmatrix}
V \\ X \\ Y \\ z_1 \\ z_2 \\ z_3 
\end{pmatrix}\leftrightarrow 
\left(\left.\begin{array}{ccccc}
-2 &-3 &0 & 0 \\
1 & 0 & 0 & 0 \\
0 & 1 & 0 & 0 \\
-2 &-3 & 1 & 0  \\
-2 &-3 & 0 & 1  \\
-2 &-3 & -1&-1 
\end{array}\right| 
\begin{array}{cc}
1 & 0 \\
2 & 6 \\
3 & 9 \\
0 & 1 \\
0 & 1 \\
0 & 1 \\
\end{array}\right)
\leftrightarrow\begin{pmatrix}\sigma_{loc} \\
2\sigma_{loc}+6H\\
3\sigma_{loc}+9H\\
H\\
H\\
H\\
\end{pmatrix}
\end{equation}
and Stanley-Reisner ideal
\begin{equation}
SRI=\{VXY,z_1z_2z_3\}\,.
\end{equation}

Let us do a simple calculation with
\begin{equation}\lambda = \frac{1}{2}\qquad \CL_{\mathbf{10}}=\CO_{\CX_4}\left(5\sigma_{loc}+6H\right)\end{equation}
If we write $\Sigma_{\mathbf{10}}$ as a complete intersection inside $\CX_4$ of
\begin{equation}\begin{split}Y^2 &= X^3 + c_0 V^6 + c_2 V^4X + c_3 V^3Y + c_4 V^2X^2 + c_5 VXY \\
0 &= b_5XY+b_4VX^2+b_3V^2Y+b_2V^3X+b_0V^5 \\
0 &= V
\end{split}\end{equation}
then \texttt{cohomcalg} \cite{Blumenhagen:2010pv, cohomCalg:Implementation} immediately gives 
\be
h^m(\Sigma_{\mathbf{10}},\CL_{\mathbf{10}})=(0,38)\,.
\ee
{This is the result of computing the chirality purely from a global perspective. }


\subsection{Computation in case of  Split of $\CC_{Higgs}$}

{We will next compare the global computation in the last subsection to the case when the local Higgs bundle spectral cover splits, and show that the correct chirality is only accounted for when non-trivial gluing is included, which is encoded in the restriction of $\mathcal{L}_{loc}$ to the Higgs bundle spectral cover $\CC_{Higgs}$.}

Consider the instance, when  the $b_m$'s are chosen so that $\CC_{Higgs}$ exhibits a 3+2 split.  In this case, we will have a factorization for the section $b_5$
\begin{equation}b_5 = a_3e_2 \,.
\end{equation}
One simple possibility is for our quartic, $b_5$, to split a pair of quadrics.

\subsubsection{Split $\CC_{Higgs}$}

To construct the 3+2 split Higgs bundle spectral cover $\CC_{Higgs}$, note that from  \cite{Dolan:2011iu}, we know that the  sections $b_m$ take the form, where $a_n$ and $e_n$ are the coefficients in the 3 and 2 factors, respectively
\begin{equation}\begin{split}b_0 &= -a_1\alpha e_0^2 \\
b_2 &= a_1\alpha\beta^2+a_2\alpha e_0 - a_1e_0e_2 \\
b_3 &= a_2\alpha\beta+a_3\alpha e_0 + a_1\beta e_2 \\
b_4 &= a_3\alpha\beta + a_2e_2 \\
b_5 &= a_3e_2
\end{split}\end{equation}
so that $\CC_{Higgs}$ is defined by
\begin{equation}0=b_0s^5+b_2s^3+b_3s^2+b_4s+b_5 = \left(a_3+a_2s + a_1s^2[\beta - e_0 s]\right)\left(e_2+\alpha s [\beta + e_0 s]\right)\,.\end{equation}
The coefficients are sections of the bundles
\begin{equation}\begin{array}{c|c}
\text{Section} & \text{Bundle} \\ \hline
a_3 & \eta - 5c_1 - \xi_A-\xi_B \\
a_2 & \eta - 4c_1 - \xi_A-\xi_B \\
a_1 & \eta - 3c_1 - \xi_A - 2\xi_B \\
e_2 & \xi_A+\xi_B \\
e_0 & c_1+\xi_B \\
\alpha & c_1 + \xi_A \\
\beta & \xi_B
\end{array}\end{equation}
For simplicity, let's choose $\xi_B = \CO$.  Further, we take $\xi_A=\CO(2H)$ so that
\begin{equation}\begin{array}{c|c}
\text{Section} & \text{Bundle} \\ \hline
a_3 & \CO(2H) \\
a_2 & \CO(5H) \\
a_1 & \CO(8H) \\
e_2 & \CO(2H) \\
e_0 & \CO(3H) \\
\alpha & \CO(5H) \\
\beta & \CO
\end{array}\end{equation}


\subsubsection{Chirality Computation in $\CC_{Higgs}$ with Gluing}

When $b_5$ splits as
\begin{equation}b_5 = a_3 e_2\end{equation}
then $\Sigma_{\mathbf{10}}$ is reducible into components
\begin{equation}\Sigma_{\mathbf{10}}\rightarrow\Sigma_{\mathbf{10}}^{(3)}\cup\Sigma_{\mathbf{10}}^{(2)} \,,
\end{equation}
where we can realize each component as a complete intersection inside $Z_3$
\begin{equation}\begin{split}\Sigma_{\mathbf{10}}^{(3)} &:\quad a_3=v=0 \\
\Sigma_{\mathbf{10}}^{(2)} &:\quad e_2=v=0
\end{split}\end{equation}
and hence as a complete intersection inside $\CX_4$, as defined in section \ref{subsec:Cloc}.  We choose $a_3$ and $e_2$ to be quadrics, so that each of these curves is a $\mathbb{P}^1$.  This means we can compute individual cohomologies of $\CL_{\mathbf{10}}|_{\Sigma_{\mathbf{10}}^{(a)}}$ uniquely from the degree.  In particular, we have that
\begin{equation}\begin{split}\CL_{\mathbf{10}}|_{\Sigma_{\mathbf{10}}^{(3)}} &= \CO_{\mathbb{P}^1}(-18)\\
\CL_{\mathbf{10}}|_{\Sigma_{\mathbf{10}}^{(2)}} &= \CO_{\mathbb{P}^1}(-18)
\end{split}\end{equation}
so that
\begin{equation}\begin{split}h^m(\Sigma_{\mathbf{10}}^{(3)},\CL_{\Sigma_{\mathbf{10}}^{(3)}})&= (0,17) \\
h^m(\Sigma_{\mathbf{10}}^{(2)},\CL_{\Sigma_{\mathbf{10}}^{(2)}})&= (0,17) \,.
\end{split}\end{equation}
When we treat the bundles separately we find a different net spectrum and indeed a different net index!  This is not unexpected: 
this  is an example how chirality of spectrum
jumps not due to gluing data but due to ramification divisors being different for the total curve and for the two parts of the curve, {corresponding to the case (1) in the Introduction. 
Indeed, 
\begin{equation}\begin{split}\chi(\Sigma_{\mathbf{10}},\CO_{\Sigma_{\mathbf{10}}}) &= 1-g_{\Sigma_{\mathbf{10}}} \\
&= -2 \\
\chi(\Sigma_{\mathbf{10}}^{(3)},\CO_{\Sigma_{\mathbf{10}}^{(3)}}) &= 1-g_{\Sigma_{\mathbf{10}}^{(3)}}\\
&= 1 \\
\chi(\Sigma_{\mathbf{10}}^{(3)},\CO_{\Sigma_{\mathbf{10}}^{(2)}}) &= 1-g_{\Sigma_{\mathbf{10}}^{(2)}}\\
&= 1 \\
\end{split}\end{equation}
so the index of the structure sheaf on $\Sigma_{\mathbf{10}}$ is different from the sums of the indices on the components $\Sigma_{\mathbf{10}}^{(a)}$.

If we are to treat the components as part of a single degenerate curve with sheaf obtained by restricting $\CN_{loc}$, the proper computation comes from the sequence
\begin{equation}\begin{split}
0 &\rightarrow H^0(\Sigma_{\mathbf{10}}^{(b)},\CL_{\mathbf{10}}(-\Sigma_{\mathbf{10}}^{(a)})|_{\Sigma_{\mathbf{10}}^{(b)}}) \rightarrow {H^0(\Sigma_{\mathbf{10}},\CL_{\mathbf{10}})}
\rightarrow H^0(\Sigma_{\mathbf{10}}^{(a)},\CL_{\mathbf{10}}|_{\Sigma_{\mathbf{10}}^{(a)}}) \\
&\rightarrow  H^1(\Sigma_{\mathbf{10}}^{(b)},\CL_{\mathbf{10}}(-\Sigma_{\mathbf{10}}^{(a)})|_{\Sigma_{\mathbf{10}}^{(b)}}) \rightarrow {H^1(\Sigma_{\mathbf{10}},\CL_{\mathbf{10}})}
\rightarrow H^1(\Sigma_{\mathbf{10}}^{(a)},\CL_{\mathbf{10}}|_{\Sigma_{\mathbf{10}}^{(a)}}) \rightarrow 0 \,.
\end{split}\label{seq1}\end{equation}
To compute this, we need to specify the map from 
\be
H^0(\Sigma_{\mathbf{10}}^{(a)},\CL_{\mathbf{10}}|_{\Sigma_{\mathbf{10}}^{(a)}})\rightarrow H^1(\Sigma_{\mathbf{10}}^{(b)},\CL_{\mathbf{10}}(-\Sigma_{\mathbf{10}}^{(a)})|_{\Sigma_{\mathbf{10}}^{(b)}})\,,
\ee
which is possible by using the exact sequence
\begin{equation}\begin{split}
0 & \rightarrow H^0(\Sigma_{\mathbf{10}}^{(b)},\CL_{\mathbf{10}}(-\Sigma_{\mathbf{10}}^{(a)})|_{\Sigma_{\mathbf{10}}^{(b)}})\rightarrow H^0(\Sigma_{\mathbf{10}}^{(b)},\CL_{\mathbf{10}}|_{\Sigma_{\mathbf{10}}^{(b)}})\rightarrow H^0(\Sigma_{\mathbf{10}}^{(a)}\cap\Sigma_{\mathbf{10}}^{(b)},\CL_{\mathbf{10}}|_{\Sigma_{\mathbf{10}}^{(a)}\cap\Sigma_{\mathbf{10}}^{(b)}}) \\
&\rightarrow H^1(\Sigma_{\mathbf{10}}^{(b)},\CL_{\mathbf{10}}(-\Sigma_{\mathbf{10}}^{(a)})|_{\Sigma_{\mathbf{10}}^{(b)}}) \rightarrow H^1(\Sigma_{\mathbf{10}}^{(b)},\CL_{\mathbf{10}}|_{\Sigma_{\mathbf{10}}^{(b)}})\rightarrow 0
\end{split}\label{seq2}\end{equation}
We take sections of $\CL_{\mathbf{10}}|_{\Sigma_{\mathbf{10}}^{(a)}}$ on $\Sigma_{\mathbf{10}}^{(a)}$ and restrict them to the intersection to get elements of $H^0(\Sigma_{\mathbf{10}}^{(a)}\cap\Sigma_{\mathbf{10}}^{(b)},\CL_{\mathbf{10}}|_{\Sigma_{\mathbf{10}}^{(a)}\cap\Sigma_{\mathbf{10}}^{(b)}})$ and then apply the coboundary map of the above sequence.  In our case, the component curves are $\mathbb{P}^1$'s and the number of intersection points is just 4, so that 
\begin{equation}\#\Sigma_{\mathbf{10}}^{(a)}\cap \Sigma_{\mathbf{10}}^{(b)}=4\,.
\end{equation}
Therefore the sequence \eqref{seq1} becomes
\begin{equation}\begin{split}0&\rightarrow H^0(\mathbb{P}^{1,(b)},\CO_{\mathbb{P}^1}(-22))\rightarrow 
{H^0(\Sigma_{\mathbf{10}},\CL_{\mathbf{10}})}\rightarrow H^0(\mathbb{P}^{1,(a)},\CO_{\mathbb{P}^1}(-18)) \\
&\rightarrow H^1(\mathbb{P}^{1,(b)},\CO_{\mathbb{P}^1}(-22))\rightarrow 
{H^1(\Sigma_{\mathbf{10}},\CL_{\mathbf{10}})}\rightarrow H^1(\mathbb{P}^{1,(a)},\CO_{\mathbb{P}^1}(-18)) \,.
\end{split}\end{equation}
We don't need to know any details of the coboundary map to conclude
\begin{equation}h^m(\Sigma_{\mathbf{10}},\CL_{\mathbf{10}}|_{\Sigma_{\mathbf{10}}})=(0,38)\end{equation}
in agreement with our previous computation.

So, if we were in a situation where the factorization property of $\CC_{Higgs}$ were global and we could talk about the two components $\Sigma_{\mathbf{10}}^{(a)}$ as distinct with each housing the bundle $\CN_{loc}|_{\Sigma_{\mathbf{10}}^{(a)}}$ we would find 17 $\mathbf{\overline{10}}$'s on each component for a total of 34 $\mathbf{\overline{10}}$'s and no $\mathbf{10}$'s.  If the factorization structure of $\CC_{Higgs}$ does not persist, however, and $\CC_{loc}$ (and indeed $\CC_F$) are irreducible so that the components are really part of a degenerate matter curve housing a sheaf obtained by restricting $\CN_{loc}$, complete with 'gluing', we get 38 $\mathbf{\overline{10}}$'s and 0 $\mathbf{10}$'s.  The extra 4 are associated to the 4 points of intersection where the degenerate curves come together.

\section{Gluing when $\CC_F$ Splits}

Throughout this discussion we have assumed that $\CC_F$ is smooth because it is the generic situation.  When $\CC_F$ is not smooth, however, we must be careful how we specify $\CN_F$.  Perhaps the most interesting situation is when $\CC_F$ splits into multiple components.  Whenever this happens, it indicates an extra $U(1)$ in the problem.

\subsection{Gluing and larger gauge group on $S_2$}

One situation in which $\CC_F$ splits into components can arise when the gauge group at $z=0$ is actually bigger than $SU(5)$.  When it is $SU(6)$, for instance, the $\bh_m$'s are such that 
\begin{equation}\bh_0z^5 + \bh_2z^3x + \bh_3 z^2y + \bh_4zx^2 + \bh_5xy = \left(\delta_0 z^3 + \delta_2 zx + \delta_3 y\right)\left(\rho_0 z^2 + \rho_2 y\right)\end{equation}
The case of $SO(10)$, on the other hand, corresponds to $\bh_5$ identically zero in which case $\CC_F$ becomes
\begin{equation}z\left(\bh_0 z^4 + \bh_2 z^2x + \bh_3 zy + \bh_4 x^2\right)\end{equation}
Models with $E_6$ gauge group have $\bh_5=\bh_4=0$ so $\CC_F$ has three components since the $z=0$ piece picks up multiplicity two
\begin{equation}\label{E6SC}
z^2\left(\bh_0z^3 + \bh_2 zx + \bh_3y\right)\,.
\end{equation}

In each of these cases, a splitting of $\CC_{Higgs}$ extends to a splitting of $\CC_F$ so we cannot fix the ambiguity in $\CN_{Higgs}$ by appealing to some well-defined line bundle on $\CC_F$.  Rather, the singular nature of $\CC_F$ means that we should be able to specify a sheaf $\CN_F$ that can be described in terms of line bundles on the individual components and gluing data.  So, the gluing data of the local model is specified by a gluing data on $\CC_F$.  But to what does gluing data on $\CC_F$ correspond?

Let's take the $SU(6)$ example to be specific and refer to the cubic and quadratic components of $\CC_F$ by $\CC_F^{(3)}$ and $\CC_F^{(2)}$, respectively.  Suppose now we consider the line bundle $\CN_{inherited}$ on the ambient 5-fold whose restriction to a smooth $\CC_F$ would yield the universal 'inherited' flux.   We can construct two different $\CN_F$'s on our degenerate $\CC_F$ from this.  One is the pair of line bundles that we get by restricting $\CN_{inherited}$ individually to the two components of $\CC_F$.  The other is the sheaf that we get by direct restriction of $\CN_{inherited}$ to $\CC_F$, which will add nontrivial gluing data along the intersection of $\CC_F^{(3)}$ and $\CC_F^{(2)}$.  {We think that some kind of non-abelian 3-form
potential $C_3$ with non-zero off-diagonal term should be appropriate description of the non-trivial global gluing.}
{This is supported by considering the local limit, where along the $z^2=0$ factor of (\ref{E6SC}) results in a degenerate spectral cover with a sheaf, that is isomorphic to the heterotic data specifying a rank 2 bundle on the heterotic side\footnote{In the heterotic context, this has been considered in \cite{Aspinwall:1998he}.}.}



\subsection{Extra $U(1)$'s not on $S_2$}

The spectral divisor $\CC_F$ also splits when extra $U(1)$'s not associated to the gauge group on $S_2$ appear.  Let us consider the example of a 4+1 split
situation that arises is when $\CC_F$ splits into two components as in
\begin{equation}\CC_F = \CC_F^{(4)} + \CC_F^{(1)}\end{equation}
In this case, the divisor $4\CC_F^{(1)}-\CC_F^{(4)}$ determines, up to some correction terms, a $(1,1)$-form that yields a new $U(1)$ symmetry (in the previous example this $(1,1)$-form would be one of the Cartans of the larger gauge group).  A pure $U(1)$ flux can be engineered by introducing separate line bundles $\CN_F^{(4)}$ and $\CN_F^{(1)}$ on the two components.  We can also add gluing data, however, it is not clear
if there is any interpretation in terms of non-abelian $C_3$ in this case.


A situation like this can occur, for example, when the $\hat{b}_m$'s are as in \eqref{bhfactor} and $\ch_m=0$ for all $m$ in \eqref{Y4def}.  In that case, the auxiliary 3-fold $Z_3$ is actually singular
\begin{equation}Wq^2 = u^2 X^3\end{equation}
Under the identification \eqref{replacement} this becomes
\begin{equation}Y^2 = X^3\end{equation}
and the equation \eqref{Clochet} for $\CC_{loc}$ can be written as
\begin{equation}\begin{split}0 &= b_0 V^5 + b_2 V^3 X + b_3 V^2 Y + b_4VX^2 + b_5 XY \\
&= \left(a_4 t^4 + a_3 t^3 v + a_2 t^2v^2 + \alpha v^3 [e_1 t - e_0 v]\right)\left(e_0 v + e_1 t\right)
\end{split}\end{equation}
where we made the replacement $X=t^2$, $Y=t^3$.  The two components $\CC_{loc}^{(4)}$ and $\CC_{loc}^{(1)}$ are the restrictions of $\CC_F^{(4)}$ and $\CC_F^{(1)}$ and if we build $G$-fluxes from separate line bundles $\CN_F^{(m)}$ on $\CC_F^{(m)}$ then the local model bundles $\CN_{loc}^{(m)}$ will be restrictions of those.  This freedom to introduce separate bundles on the two components of $\CC_F$ represents the possibility of turning on $U(1)$ flux.

Note that the splitting of $\CC_{loc}$ in this case was made possible by the degeneration of the auxiliary 3-fold $Z_3$ in which it is naturally embedded.  More specifically, the elliptic fiber of $Z_3$ was pinched everywhere to a $\mathbb{P}^1$ fiber.  If we ignore the fact that this $\mathbb{P}^1$ is really a pinched torus it is easy to see that this 3-fold is just $\mathbb{P}(\CO_{S_2}\oplus K_{S_2})$, namely the usual one used to study compactifications of $\CC_{Higgs}$ in local model studies.  The factorization structure of $\CC_{Higgs}$ remains in this case because this particular compactification, which preserves it, is the actual one that we find in $Y_4$.


\subsection*{Acknowledgements}

We thank Herb Clemens, Mirjam Cvetic, Ron Donagi, Jim Halverson and Moritz Kuntzler for helpful discussions.
SSN thanks the Bethe Center in Bonn and the IFT Madrid for hospitality during the completion of this work. 
JM is supported by DOE grant DE-FG02-90ER-40560 and NSF grant PHY-0855039, and SSN is supported in part by STFC.

\newpage



\begin{thebibliography}{10}

\bibitem{Donagi:2009ra}
R.~Donagi and M.~Wijnholt, {\it {Higgs Bundles and UV Completion in F-Theory}},
   \href{http://xxx.lanl.gov/abs/0904.1218}{{\tt 0904.1218}}.

\bibitem{Cecotti:2010bp}
S.~Cecotti, C.~Cordova, J.~J. Heckman, and C.~Vafa, {\it {T-Branes and
  Monodromy}},  \href{http://xxx.lanl.gov/abs/1010.5780}{{\tt 1010.5780}}.

\bibitem{Donagi:2011jy}
R.~Donagi and M.~Wijnholt, {\it {Gluing Branes, I}},
  \href{http://xxx.lanl.gov/abs/1104.2610}{{\tt 1104.2610}}.

\bibitem{Donagi:2011dv}
R.~Donagi and M.~Wijnholt, {\it {Gluing Branes II: Flavour Physics and String
  Duality}},  \href{http://xxx.lanl.gov/abs/1112.4854}{{\tt 1112.4854}}.

\bibitem{Tatar:2009jk}
R.~Tatar, Y.~Tsuchiya, and T.~Watari, {\it {Right-handed Neutrinos in F-theory
  Compactifications}},  {\em Nucl.Phys.} {\bf B823} (2009) 1--46,
  [\href{http://xxx.lanl.gov/abs/0905.2289}{{\tt 0905.2289}}].

\bibitem{Marsano:2009gv}
J.~Marsano, N.~Saulina, and S.~Schafer-Nameki, {\it {Monodromies, Fluxes, and
  Compact Three-Generation F-theory GUTs}},  {\em JHEP} {\bf 08} (2009) 046,
  [\href{http://xxx.lanl.gov/abs/0906.4672}{{\tt 0906.4672}}].

\bibitem{Marsano:2009wr}
J.~Marsano, N.~Saulina, and S.~Schafer-Nameki, {\it {Compact F-theory GUTs with
  $U(1)_{PQ}$}},  {\em JHEP} {\bf 04} (2010) 095,
  [\href{http://xxx.lanl.gov/abs/0912.0272}{{\tt 0912.0272}}].

\bibitem{Hayashi:2010zp}
H.~Hayashi, T.~Kawano, Y.~Tsuchiya, and T.~Watari, {\it {More on Dimension-4
  Proton Decay Problem in F-theory -- Spectral Surface, Discriminant Locus and
  Monodromy}},  {\em Nucl.Phys.} {\bf B840} (2010) 304--348,
  [\href{http://xxx.lanl.gov/abs/1004.3870}{{\tt 1004.3870}}].

\bibitem{Marsano:2010ix}
J.~Marsano, N.~Saulina, and S.~Schafer-Nameki, {\it {A Note on G-Fluxes for
  F-theory Model Building}},  {\em JHEP} {\bf 11} (2010) 088,
  [\href{http://xxx.lanl.gov/abs/1006.0483}{{\tt 1006.0483}}].

\bibitem{Dudas:2010zb}
E.~Dudas and E.~Palti, {\it {On hypercharge flux and exotics in F-theory
  GUTs}},  {\em JHEP} {\bf 09} (2010) 013,
  [\href{http://xxx.lanl.gov/abs/1007.1297}{{\tt 1007.1297}}].

\bibitem{Grimm:2010ez}
T.~W. Grimm and T.~Weigand, {\it {On Abelian Gauge Symmetries and Proton Decay
  in Global F-theory GUTs}},  {\em Phys.Rev.} {\bf D82} (2010) 086009,
  [\href{http://xxx.lanl.gov/abs/1006.0226}{{\tt 1006.0226}}].

\bibitem{Marsano:2011nn}
J.~Marsano, N.~Saulina, and S.~Schafer-Nameki, {\it {On G-flux, M5 instantons,
  and U(1)s in F-theory}},  \href{http://xxx.lanl.gov/abs/1107.1718}{{\tt
  1107.1718}}.

\bibitem{Dolan:2011iu}
M.~J. Dolan, J.~Marsano, N.~Saulina, and S.~Schafer-Nameki, {\it {F-theory GUTs
  with U(1) Symmetries: Generalities and Survey}},
  \href{http://xxx.lanl.gov/abs/1102.0290}{{\tt 1102.0290}}.

\bibitem{MS}
J.~Marsano and S.~Schafer-Nameki, {\it {Yukawas, G-flux, and Spectral Covers
  from Resolved Calabi-Yau's}},  {\em JHEP} {\bf 1111} (2011) 098,
  [\href{http://xxx.lanl.gov/abs/1108.1794}{{\tt 1108.1794}}].

\bibitem{Kuntzler:2012bu}
M.~Kuntzler and S.~Schafer-Nameki, {\it {G-flux and Spectral Divisors}},
  \href{http://xxx.lanl.gov/abs/1205.5688}{{\tt 1205.5688}}.

\bibitem{Collinucci:2010gz}
A.~Collinucci and R.~Savelli, {\it {On Flux Quantization in F-Theory}},
  \href{http://xxx.lanl.gov/abs/1011.6388}{{\tt 1011.6388}}.

\bibitem{Braun:2011zm}
A.~P. Braun, A.~Collinucci, and R.~Valandro, {\it {G-flux in F-theory and
  algebraic cycles}},  {\em Nucl.Phys.} {\bf B856} (2012) 129--179,
  [\href{http://xxx.lanl.gov/abs/1107.5337}{{\tt 1107.5337}}]. 55 pages, 1
  figure/ added refs, corrected typos.

\bibitem{Krause:2011xj}
S.~Krause, C.~Mayrhofer, and T.~Weigand, {\it {$G_4$ flux, chiral matter and
  singularity resolution in F-theory compactifications}},  {\em Nucl.Phys.}
  {\bf B858} (2012) 1--47, [\href{http://xxx.lanl.gov/abs/1109.3454}{{\tt
  1109.3454}}].

\bibitem{Krause:2012yh}
S.~Krause, C.~Mayrhofer, and T.~Weigand, {\it {Gauge Fluxes in F-theory and
  Type IIB Orientifolds}},  {\em JHEP} {\bf 1208} (2012) 119,
  [\href{http://xxx.lanl.gov/abs/1202.3138}{{\tt 1202.3138}}].

\bibitem{Collinucci:2012as}
A.~Collinucci and R.~Savelli, {\it {On Flux Quantization in F-Theory II:
  Unitary and Symplectic Gauge Groups}},  {\em JHEP} {\bf 1208} (2012) 094,
  [\href{http://xxx.lanl.gov/abs/1203.4542}{{\tt 1203.4542}}].

\bibitem{Cvetic:2012ts}
M.~Cvetic, R.~Donagi, J.~Halverson, and J.~Marsano, {\it {On Seven-Brane
  Dependent Instanton Prefactors in F-theory}},
  \href{http://xxx.lanl.gov/abs/1209.4906}{{\tt 1209.4906}}.

\bibitem{Blumenhagen:2010pv}
R.~Blumenhagen, B.~Jurke, T.~Rahn, and H.~Roschy, {\it {Cohomology of Line
  Bundles: A Computational Algorithm}},  {\em J. Math. Phys.} {\bf 51} (2010)
  103525, [\href{http://xxx.lanl.gov/abs/1003.5217}{{\tt 1003.5217}}].

\bibitem{cohomCalg:Implementation}
``{cohomCalg package}.'' Download link, 2010.
\newblock High-performance line bundle cohomology computation based on
  \cite{Blumenhagen:2010pv}.

\bibitem{Aspinwall:1998he}
P.~S. Aspinwall and R.~Y. Donagi, {\it {The Heterotic string, the tangent
  bundle, and derived categories}},  {\em Adv.Theor.Math.Phys.} {\bf 2} (1998)
  1041--1074, [\href{http://xxx.lanl.gov/abs/hep-th/9806094}{{\tt
  hep-th/9806094}}].

\end{thebibliography}

\providecommand{\href}[2]{#2}\begingroup\raggedright\endgroup


\end{document}